\documentstyle[prl,aps,epsfig,amsmath,twocolumn]{revtex}
\begin{document}
\input epsf
\twocolumn[\hsize\textwidth\columnwidth\hsize\csname  
@twocolumnfalse\endcsname
\title{Subharmonic Shapiro steps and assisted tunneling in superconducting 
point contacts}

\author{J.C. Cuevas$^1$, J. Heurich$^1$, A. Mart\'{\i}n-Rodero$^2$,
A. Levy Yeyati$^2$ and G. Sch\"on$^{1,3}$}

\address{$^1$ Institut f\"ur Theoretische Festk\"orperphysik, Universit\"at
Karlsruhe, 76128 Karlsruhe, Germany \\
$^2$ Departamento de F\'\i sica Te\'orica de la Materia Condensada
C-V, Universidad Aut\'onoma de Madrid, E-28049 Madrid, Spain \\
$^3$ Forschungszentrum Karlsruhe, Institut f\"ur Nanotechnologie, D-76021 Karlsruhe, Germany }


\maketitle

\begin{abstract}
We analyze the current in a superconducting point contact of arbitrary transmission 
in the presence of a microwave radiation. The interplay between the ac Josephson 
current and the microwave signal gives rise to Shapiro steps at voltages 
$V = (m/n) \hbar \omega_r/2e$, where $n,m$ are integer numbers and $\omega_r$ is 
the radiation frequency. The \emph{subharmonic steps} ($n \ne 1$) are a 
consequence of multiple Andreev reflections (MAR) and provide 
a signature of the peculiar ac Josephson effect at high transmission. Moreover, 
the dc current exhibits a rich subgap structure due to photon-assisted MARs.
\end{abstract}


\vskip2pc]
\narrowtext

{\em Introduction.}--- 
Our understanding of the electronic transport through superconducting 
nanostructures has experienced a notable development in last few years \cite{SM1999}. 
Partly, this has been due to the appearance on scene of the metallic 
atomic-size contacts, which can be produced by means of scanning tunneling
microscope and break-junction techniques \cite{Muller1992,Scheer1997,Goffman2000,Cron2001}. 
These nanowires have turned out to be ideal systems to test the modern transport theories  
in mesoscopic superconductors. Thus, for instance Scheer and coworkers 
\cite{Scheer1997} found a quantitative agreement between the measurements of the 
current-voltage characteristics of different atomic contacts and the predictions of the 
theory for a single-channel superconducting contact \cite{Bratus1995,Cuevas1996}. 
These experiments not only helped to clarify the structure of the subgap current in
superconducting contacts, but also showed that the set of the transmission coefficients 
in an atomic-size contact is amenable to measurement. This possibility has recently 
allowed a set of experiments that confirm the theoretical predictions for transport 
properties like supercurrent \cite{Goffman2000} and noise \cite{Cron2001}. 
From these combined theoretical and experimental efforts a coherent picture of
transport in superconducting point contacts has emerged with multiple Andreev
reflections (MAR) \cite{Klapwijk1982} as a central concept. However,
in spite of these recent successes, one of the most remarkable predictions of MAR 
theory remains to be confirmed, namely the \emph{ac Josephson effect}. The theory says 
that in a constant voltage biased superconducting point contact, the time-dependent current 
is given by $I(t) = \sum_n I_n e^{in \omega_0 t}$. This means that the occurrence 
of MARs gives rise to the appearance of alternating currents that
oscillate not only with the Josephson frequency $\omega_0=2eV/\hbar$, $V$ being 
the voltage, as in the case of tunnel junctions, but also with all its harmonics. 
So far there is no experimental evidence of the existence of such components. 

In this paper, we present a theoretical analysis of the current 
in a superconducting point contact under a microwave radiation. We show that the 
interplay between the ac Josephson current components and a microwave signal leads to the 
appearance of Shapiro steps at voltages $V = (m/n) \hbar \omega_r/2e$, where $n,m$ are 
integer numbers and $\omega_r$ is the frequency of the radiation. This means that in addition 
to the usual steps ($n=1$) found in tunnel junctions \cite{Shapiro1963}, there also appear 
\emph{subharmonic Shapiro steps} ($n \ne 1$), which constitute an unambiguous 
signature of the ac Josephson effect in these contacts. Moreover, we also find that the dc 
background current, in which the Shapiro steps are superimposed, exhibits a rich subgap 
structure, which can be understood in terms of photon-assisted MARs and provides a natural 
explanation of experimental findings in the early seventies \cite{SGS}.

{\em Theoretical model.}--- 
Our goal is to calculate the current in a voltage biased superconducting 
quantum point contact (SQPC) \cite{note1} in the presence of a monochromatic
radiation of frequency $\omega_r$. We assume that the external radiation
produces an effective time-dependent voltage $V(t) = V + V_{ac} \sin \omega_r t$. Our task
is to extend the MAR theory to the case of such a time-dependent voltage, for which the
so-called Hamiltonian approach \cite{Cuevas1996} is a convenient starting point. 
For the voltage range $eV \sim \Delta$ one can neglect the energy dependence of the 
transmission coefficients and all transport properties can be expressed as a superposition 
of independent channel contributions. Thus, the problem reduces to the analysis of a single
channel contact, which can be described by means of the following tight-binding-like 
Hamiltonian \cite{Cuevas1996}

\begin{equation}
\hat{H} = \hat{H}_{L} + \hat{H}_{R} + 
\sum_{\sigma} \left\{ v \; c^{\dagger}_{L \sigma} c_{R \sigma}
+ v^{*} \; c^{\dagger}_{R \sigma} c_{L \sigma} \right\}  ,
\end{equation}

\noindent
where $H_{L,R}$ are the BCS Hamiltonians for the isolated electrodes. In the
coupling term $L$ and $R$ stand for the outermost sites of each electrode, and
$v$ is a hopping parameter coupling these sites. This parameter determines 
the normal transmission coefficient of this model ${\cal{T}}$, which adopts
the form ${\cal{T}} = 4(v/W)^2/ \left[1+(v/W)^2 \right]^2$, where $W=1/\pi\rho_F$,
with $\rho_F$ being the electrodes density of states at the Fermi energy 
\cite{Cuevas1996}.

In this model the current evaluated at the interface between the two electrodes
adopts the form

\begin{equation}
\hspace*{-3mm} I(t) = \frac{i e}{\hbar}  \sum_{\sigma} \left\{ v \langle 
c^{\dagger}_{L \sigma}(t) c_{R \sigma}(t) \rangle - 
v^{*} \langle c^{\dagger}_{R \sigma}(t) c_{L \sigma}(t) \rangle \right\} .
\label{current}
\end{equation}

\noindent
The non-equilibrium expectation values in Eq. (\ref{current}) can be expressed in 
terms of the Keldysh Green functions $\hat{G}^{+,-}_{i,j}$, 
which in the $2 \times 2$ Nambu representation read

\begin{equation}
\hat{G}^{+-}_{i,j}(t,t^{\prime})= i \left( \begin{array}{cc}
\langle c^{\dagger}_{j \uparrow} (t^{\prime}) c_{i \uparrow}(t) \rangle   &
\langle c_{j \downarrow}(t^{\prime}) c_{i \uparrow}(t) \rangle \\
\langle c^{\dagger}_{j \uparrow}(t^{\prime}) c^{\dagger}_{i \downarrow}(t) 
\rangle  & \langle c_{j \downarrow}(t^{\prime}) c^{\dagger}_{i \downarrow}(t)
\rangle \end{array}  \right) .
\end{equation}

Thus, the current can be now written as

\begin{equation}
I(t)  =  \frac{e}{\hbar} \; \mbox{Tr} \left[ \hat{\tau}_3
\left( \hat{v}(t) \hat{G}^{+-}_{RL}(t,t) -
\hat{v}^{\dagger}(t) \hat{G}^{+-}_{LR}(t,t) \right) \right] ,
\end{equation}

\noindent
where $\hat{\tau}_3$ is the corresponding Pauli matrix, $\mbox{Tr}$
denotes the trace in Nambu space and $\hat{v}$ is the hopping that in the
Nambu matrix representation is written as

\begin{equation}
\hat{v}(t) = \left(
\begin{array}{cc}
 v e^{i \phi(t)/2}   &   0      \\
    0                   &   -v^{*} e^{-i \phi(t)/2}
\end{array} \right)  .
\end{equation}

\noindent
Here, $\phi(t) = \phi_0 + \omega_0 t + 2 \alpha \cos{\omega_r t}$
is the time-dependent superconducting phase difference. 
The constant $\alpha = eV_{ac}/(\hbar \omega_r)$ measures the strength of the 
coupling to the electromagnetic field, and is proportional to the
square root of the radiation power. 

In order to determine the Green functions we follow a perturbative 
scheme and treat the coupling term in Hamiltonian (1) as a perturbation. The 
unperturbed Green functions, $\hat{g}$, correspond to the uncoupled electrodes 
in equilibrium. Thus, the retarded and advanced components adopt the BCS 
form: $\hat{g}^{r,a}(\epsilon) = g^{r,a}(\epsilon) \hat{1} + f^{r,a}(\epsilon) 
\hat{\tau}_1$, where $g^{r,a}(\epsilon) = - (\epsilon^{r,a} / \Delta) f(\epsilon) 
= -\epsilon^{r,a} / W \sqrt{ \Delta^{2} -(\epsilon^{r,a})^2 }$, where 
$\epsilon^{r,a} = \epsilon \pm i \eta$, with $\eta=0^+$. 
Following Ref. \cite{Cuevas1996} we express the current in terms of a 
T-matrix, rather than in terms of the Green functions. The T-matrix associated to 
the time-dependent perturbation of Eq. (5) is defined as
$\hat{T}^{r,a} = \hat{v} + \hat{v} \circ \hat{g}^{r,a} \circ \hat{T}^{r,a}$, 
where the $\circ$ product is a shorthand for integration over intermediate time 
arguments. As shown in Ref. \cite{Cuevas1996}, the current in terms of the T-matrix 
components reads

\begin{eqnarray}
\hspace*{-5mm} I(t) & = & \frac{e}{\hbar} \; \mbox{Tr} \left[ \hat{\tau}_3
\left( \hat{T}_{LR}^r \circ  \hat{g}_{R}^{+-} \circ  \hat{T}_{RL}^{a} 
\circ \hat{g}_{L}^a - \hat{g}_{L}^{r} \circ \hat{T}_{LR}^r \circ
\hat{g}_{R}^{+-} \circ  \hat{T}_{RL}^{a}
\nonumber \right. \right. \\ && \left. \left. +
\hat{g}_{R}^{r} \circ \hat{T}_{RL}^{r} \circ  \hat{g}_{L}^{+-} \circ 
\hat{T}_{LR}^{a} - \hat{T}_{RL}^{r} \circ  \hat{g}_{L}^{+-} \circ
\hat{T}_{LR}^{a} \circ \hat{g}_{R}^{a}
\right) \right] ,
\end{eqnarray}

In order to solve the T-matrix integral equation it is convenient to Fourier transform
with respect to the temporal arguments, $\hat{T}(t,t^{\prime}) = (1/2 \pi) \int d\epsilon \; 
\int d\epsilon^{\prime} \; e^{-i \epsilon t} e^{i \epsilon^{\prime} t^{\prime}}
\; \hat{T}(\epsilon,\epsilon^{\prime})$. Due to time dependence of the coupling element
(see Eq. (5)), one can show that $\hat{T}(\epsilon,\epsilon^{\prime})$ admits the following
solution: $\hat{T}(\epsilon,\epsilon^{\prime}) = \sum_{n,m} \hat{T}(\epsilon,
\epsilon + neV + m \hbar \omega_r) \delta(\epsilon - \epsilon^{\prime} + neV +
m \hbar \omega_r)$. Thus, one can finally write down the current as $I(t) = \sum_{n,m} 
I^m_n \exp \left[ i \left( n \phi_0 + n \omega_0 t + m \omega_r t \right) \right]$,
where the current amplitudes $I^m_n$ can be expressed in terms of the T-matrix Fourier 
components, $\hat{T}^{kl}_{nm} \equiv \hat{T}(\epsilon + neV + k 
\hbar \omega_r, \epsilon + meV + l \hbar \omega_r)$, in the following way

\begin{eqnarray}
I^m_n &= & \frac{e}{h} \int d\epsilon \sum_{i,k} \mbox{Tr} \left[ \hat{\tau}_3 
\times  \right.  \nonumber \\ & & \hspace{-1cm} 
\left( \hat{T}^{\substack{r \\ 0k}}_{LR,0i} \hat{g}^{\substack{+- \\ k}}_{R,i}
\hat{T}^{\substack{a \\ km}}_{RL,in} \hat{g}^{\substack{a \\m}}_{L,n}
- \hat{g}^{\substack{r \\ 0}}_{L,0} \hat{T}^{\substack{r \\ 0k}}_{LR,0i}
\hat{g}^{\substack{+- \\ k}}_{R,i} \hat{T}^{\substack{a \\ km}}_{RL,in} +
\right. \nonumber \\ & & \left. \left. \hspace{-1cm} 
\hat{g}^{\substack{r \\ 0}}_{R,0} \hat{T}^{\substack{r \\ 0k}}_{RL,0i}
\hat{g}^{\substack{+- \\ k}}_{L,i} \hat{T}^{\substack{a \\ km}}_{LR,in}
- \hat{T}^{\substack{r \\ 0k}}_{RL,0i} \hat{g}^{\substack{+- \\ k}}_{L,i}
\hat{T}^{\substack{a \\ km}}_{LR,in} \hat{g}^{\substack{a \\ m}}_{R,n} \right) 
\right].
\end{eqnarray}

At this point, the calculation of the current has been reduced to determination 
of the Fourier components of the T-matrix. In the case of a symmetric contact
considered here, one can show that the dc current can be expressed only in terms
of $ \hat{T}^k_i \equiv \hat{T}^{\substack{a \\ k0}}_{LR,i0}$, which
fulfill the following set of linear algebraic equations

\begin{equation}
\hspace*{-3mm} \hat{T}^k_i  =  \hat{v}^k_i + 
\sum_{l} \left\{  \hat{{\cal{E}}}^{kl}_{i,i} \hat{T}^l_i +
\hat{{\cal{V}}}^{kl}_{i,i+2} \hat{T}^l_{i+2} + \hat{{\cal{V}}}^{kl}_{i,i-2} 
\hat{T}^l_{i-2} \right\} ,
\end{equation}

\noindent
where the different matrix coefficients adopt the following form in terms of the
unperturbed Green functions

\begin{eqnarray}
\hat{v}^k_i & = & \frac{v}{2} J_k(\alpha_0) \left[ i^k (\hat{1} + \hat{\tau}_3
) \; \delta_{i,-1} - (-i)^k (\hat{1} - \hat{\tau}_3) \; \delta_{i,1} \right] 
\nonumber \\
\hat{{\cal{E}}}^{kl}_{i,i} & = & v^2 i^{k+l} \sum_{j} (-1)^j J_{k-j}(\alpha) 
J_{j-l}(\alpha) \left( \begin{array}{cc}
g^j_{i+1} g^l_i  & g^j_{i+1} f^l_i \\
g^j_{i-1} f^l_i  & g^j_{i-1} g^l_i
\end{array} \right) \nonumber \\
\hat{{\cal{V}}}^{kl}_{i,i+2} & = & -v^2 i^{k-l} \sum_{j} J_{k-j}(\alpha) 
J_{j-l}(\alpha) f^j_{i+1} \left( \begin{array}{cc}
f^l_{i+2} & g^l_{i+2} \\
0  & 0
\end{array} \right) \nonumber  \\
\hat{{\cal{V}}}^{kl}_{i,i-2} & = & -v^2 i^{l-k} \sum_{j} J_{k-j}(\alpha) 
J_{j-l}(\alpha) f^j_{i-1}
\left( \begin{array}{cc}
0 & 0 \\
g^l_{i-2}  & f^l_{i-2}
\end{array} \right) , \nonumber
\end{eqnarray}

\noindent
where we have used the shorthand notation $\hat{g}^k_i = \hat{g}^a(\epsilon + ieV 
+ k \hbar \omega_r)$ and $J_n(\alpha)$ is the Bessel function of order $n$. In some
limits one can find an analytical solution of these systems, but in general a numerical 
calculation is needed.

{\em Results and discussions.}--- 
Let us concentrate in the dc current, $I_{dc}$. This current is the sum of two 
contributions: $I_{dc} = I_{B} + I_{Shapiro}$, where $I_{B} \equiv I^0_0$ is a background 
current and $I_{Shapiro} = \sum_{n,m} I^m_n e^{in\phi_0} \delta(V-V^m_n)$ is the Shapiro 
steps contribution at discrete voltages $V^m_n = (m/n) \hbar \omega_r/2e$.
Notice that several ac current amplitudes can give a dc contribution at the same voltage.
Notice also that the Shapiro step contribution depends on the average value of the phase,
$\phi_0$. We shall concentrate in the height of the Shapiro steps, which will be 
denoted as $S^m_n$. Let us remark that in the tunneling regime we recover the well-known 
results for both the background current and Shapiro step heights \cite{Barone1982}.

In order to illustrate the general results, in Fig. 1 we show the dc current, background 
current plus Shapiro steps, for different values of $\alpha$ and a frequency
$\omega_r = 0.5\Delta$. We can see the two main features that
will be the subject of the rest of the paper: (i) the \emph{subharmonic Shapiro 
steps} $S^m_n$, with $n \ne 1$, are clearly visible at high transmissions, and 
(ii) the background current exhibits a subharmonic gap structure at voltages 
$eV = (2\Delta + k\hbar \omega_r)/n$, with $n,k$ integers, which is specially 
pronounced at low transmissions \cite{Gunsen98}. 

\begin{figure}[!h]
\begin{center}
\epsfig{file=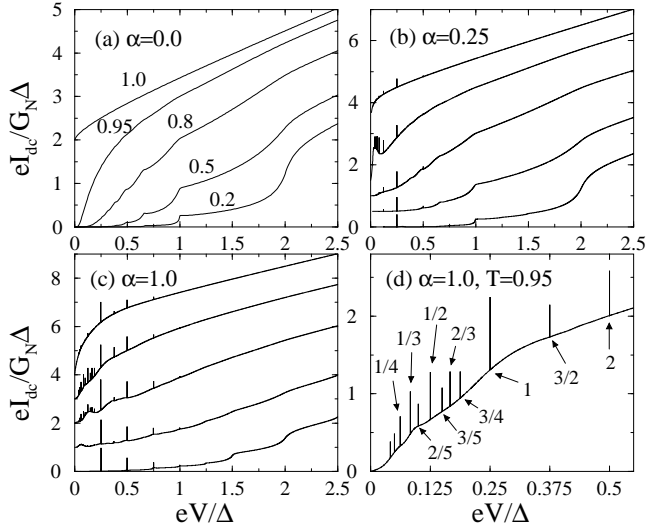,width=8.5cm,clip=,silent=}
\caption{\label{dc-current} Zero temperature dc current, $I_{dc}$,
as a function of voltage for a frequency $\omega_r=0.5\Delta$ and several values 
of $\alpha$. The different curves in each panel correspond to different transmissions,
as indicated in panel (a). In panels (b) and (c) the curves have been vertically displaced. 
Panel (d) shows in detail the curve ${\cal{T}}=0.95$ of panel (c). The current is 
normalized by the normal conductance $G_N=(2e^2/h) {\cal{T}}$.  }
\end{center}
\end{figure}

\vspace{-3mm}
Let us start by analyzing the background current.  In Fig. 2(c-d) we show the 
background current for two different frequencies at a moderate power, $\alpha=1.0$. 
The current in absence of radiation is also shown for comparison. As mentioned above, 
the most prominent feature in the background current is the appearance of a pronounced 
subgap structure at voltages $eV = (2\Delta + k\hbar \omega_r)/n$. This structure
is specially clear at low transmissions (see Fig. 2d) and progressively 
disappears as the transparency is increased. Indeed, this peculiar subharmonic gap 
structure was already observed in several experiments in the early seventies in
point contacts and thin-film microbridges \cite{SGS}. At that time no consistent 
explanation was given, but it is clear that this structure can be explained
in terms of photon-assisted MARs. A step at $eV = (2\Delta + k\omega_r)/n$ is simply due 
to the opening of a MAR of order $n$ in which $k$ photons in total are absorbed ($k$ 
negative) or emitted ($k$ positive). This is illustrated in the upper panels of Fig. 2.
In order to understand how this subharmonic structure evolves with the rf power, one
can do a systematic perturbative expansion in the transmission. This analysis tells us
that at low transparency the height of a current jump at $eV = (2\Delta + k\omega_r)/n$ 
is proportional to $J^2_k(n\alpha)$, which is valid as long as $\hbar \omega_r \ll 2\Delta/n$.
This results coincides with the phenomenological functional form that was used to fit 
the experiments by Soerensen \emph{et al.} \cite{SGS}.

\begin{figure}[!h]
\begin{center}
\epsfig{file=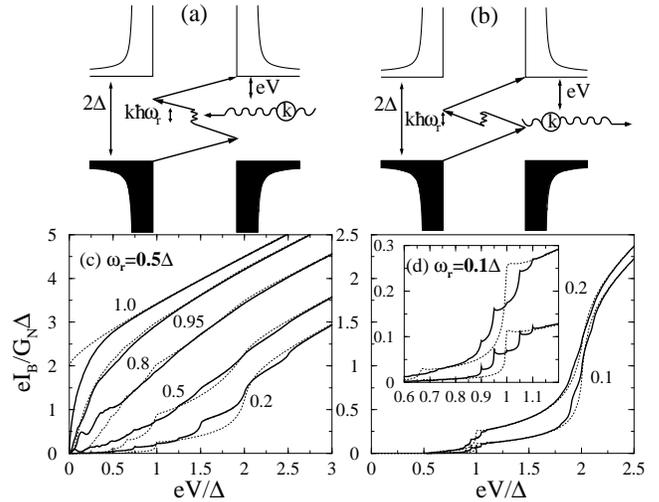,width=8.5cm,clip=,silent=}
\caption{\label{back} (a) Representation of a MAR of order $3$ in
which $k$ photons are absorbed. This process has a threshold voltage $eV_{th} = (2\Delta - 
\hbar |k|\omega_r)/3$, and its probability amplitude is proportional to $J_k(\alpha)$.
(b) A 3-order MAR mediated by the emission of $k$ photons, which contributes to the
subgap structure at $eV_{th} = (2\Delta + \hbar |k|\omega_r)/3$. (c) 
Background current as a function of voltage for $\omega_r=0.5\Delta$ and different
transmissions. (d) The same as in (c) for $\omega_r=0.1\Delta$. The inset shows a
blow up around $eV=\Delta$. The dotted lines in (c) and (d)
correspond to the current in absence of radiation.  } 
\end{center}
\end{figure}

\vspace{-3mm}
Let us now discuss the Shapiro steps. In this case the most important aspect
is the existence of subharmonic steps absent in tunnel
junctions. These steps arise from the phase locking between the harmonics of the 
Josephson frequency and the harmonics of the ac radiation. Early experiments on 
the ac Josephson effect in weak links observed subharmonic steps in the I-V curves 
\cite{Anderson1964}.  More recently, there have been reported observations 
of non-integer Shapiro steps in high-$T_C$ contacts \cite{Early1993}, S-semiconductor-S 
junctions \cite{Lehnert1999} and diffusive S-N-S systems \cite{Dubos2001}. 
Although the Shapiro steps can be understood as a simple consequence of a
non-sinusoidal current-phase relation, the present approach goes beyond a simple
``adiabatic" approximation and provides the first microscopic theory of Shapiro
steps in contacts of arbitrary transmission. The adiabatic
approximation, which introduces the time-dependence into the zero bias
supercurrent through the Josephson relation, gives rise to the well known
Bessel-function-like behavior of the steps and gives a good descprition of the
tunnel regime \cite{Barone1982}. However, as we show below, such a simple approach
fails in the description of a highly transmissive contact.

\begin{figure}[!h]
\begin{center}
\epsfig{file=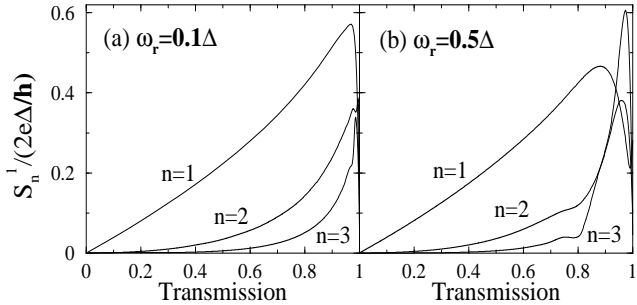,width=8.5cm,height=4.25cm,clip=,silent=}
\caption{\label{Shap-trans} Shapiro steps $S^1_n$ versus transmission for $\alpha=0.25$.}
\end{center}
\end{figure}

As a rule of thumb, a Shapiro step $S^m_n$ is visible when the corresponding ac Josephson 
component, $I_n$, in absence of radiation gives a significant contribution. In particular,
this means high transmissions (see Figs. 3-4 in Ref. \cite{Cuevas1996}).
One can show that the leading order in transmission of a Shapiro step $S^m_n$ goes
like $\sim {\cal{T}}^n$, which is a consequence of the fact that $I_n \sim {\cal{T}}^n$,
and the reason for the absence of the $n \ne 1$ steps in low transmissive contacts.
However, near perfect transmission the subharmonic steps can be even higher than the 
integer ones. This behavior is illustrated in Fig. \ref{Shap-trans}, where we
show the Shapiro steps $S^1_n$ as a function of the transmission for two different 
frequencies.

\begin{figure}[!h]
\begin{center}
\epsfig{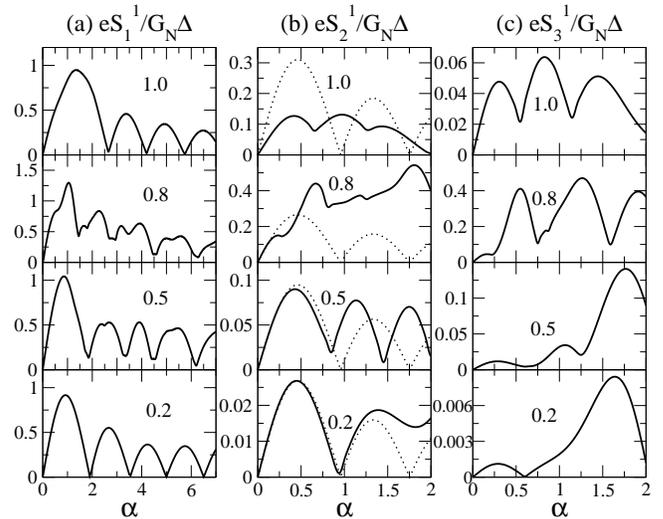}
\caption{\label{Shap-power} (a-c) Shapiro steps $S^1_n$ ($n=1,2,3$) as a function of $\alpha$
for $\omega_r=0.5\Delta$. The dotted lines in panel (b) correspond to the
adiabatic approximation: $\sim |J_1(4\alpha)|$.}
\end{center}
\end{figure}

\vspace{-3mm}
Fig. \ref{Shap-power} shows the power dependence of the Shapiro steps for a frequency
$\omega_r=0.5\Delta$. Notice that this dependence is rather complicated for both integer
and subharmonic steps, and clearly deviates from the usual Bessel function behavior. This 
is due to the frequency-dependence of the Josephson components, which is specially 
pronounced at high transmissions. Neglecting this dependence, i.e. within an
adiabatic approximation, one would get that $S^m_n$ evolves as $|J_m(2n\alpha)|$.
However, as shown in Fig. \ref{Shap-power}b, as the transmission increases the validity of 
this approximation is restricted to $\alpha \ll 1$. Notice also the complex oscillation 
pattern at high transmissions (see ${\cal{T}}=0.8$ curves in Fig \ref{Shap-power}),
which is due to the fact that several ac components give a significant contribution
to the same Shapiro step.

In summary, we have presented a theoretical analysis of the dc current in a
superconducting point contact in the presence of a microwave radiation. We have
shown that the microscopic theory of coherent multiple Andreev reflections provides
an unified description of Shapiro steps and assisted tunneling, explaining in a natural
way the observations of subharmonic steps 
\cite{Anderson1964,Early1993,Lehnert1999,Dubos2001} and the peculiar 
subharmonic gap structure under a microwave radiation \cite{SGS}. Let us finally remark 
that the results presented in this work are amenable to a quantitative experimental test
using atomic-size contacts \cite{Muller1992,Scheer1997,Goffman2000,Cron2001}.

We thank H. Courtois, R. Cron, M.F. Goffman, B. Pannetier,
E. Scheer and A. Zaikin for useful discussions. This work has been supported by the
EU TMR Network on Dynamics of Nanostructures, the CFN supported by the DFG
and the Spanish CICyT under Co. PB97-0044.

\vspace{-8mm}



\begin{references}
\bibitem{note1}
By SQPC we mean a contact between two superconducting electrodes linked by
a constriction whose length is much smaller than the superconducting coherent
length.
\end{references}
\end{document}